\documentclass[10pt,twocolumn]{revtex4}
\usepackage{amsmath}
\usepackage{latexsym,amssymb,epsfig,graphicx}

\begin{document}
\title{Molecular model for de Vries type smectic A - smectic C \\phase
transition in liquid crystals}
\author{ M.V. Gorkunov$^{1,2}$, M.A. Osipov$^1$, F. Giesselmann$^3$, T.J. Sluckin$^4$, and J.P.F. Lagerwall$^3$}
\affiliation{$^1$Department of Mathematics, University of
Strathclyde, Glasgow G1 1XH, UK \\ $^2$Institute of Crystallography,
Russian Academy of Sciences, 119333 Moscow, Russia\\ $^3$Institute
of Physical Chemistry, University of Stuttgart, 70569 Stuttgart,
Germany\\ $^4$School of Mathematics, University of Southampton,
Southampton SO17 1BJ, UK}
\begin{abstract}
We develop a theory of Smectic A - Smectic C phase transition with
anomalously weak smectic layer contraction. We construct a
phenomenological description of this transition by generalizing the
Chen-Lubensky model. Using a mean-field molecular model, we
demonstrate that a relatively simple interaction potential suffices
to describe the transition. The theoretical results are in excellent
agreement with experimental data.
\end{abstract}
\maketitle

Thermotropic smectic liquid crystals exhibit layered orientationally
ordered phases. In the simplest smectic A (Sm{\it A}) phase, the
long molecular axis director ${\bf n}$ is normal to the smectic
layers and the phase is uniaxial \cite{DeGennes,Lagerwall}. But in
the smectic C (Sm{\it C}) phase, the director is inclined at an
angle $\Theta$ to the layer normal, and the phase is biaxial. More
complex smectic order has also been observed \cite{revFrankJan}.
Analogous tilted smectic order is observed in other soft-matter
systems, such as block copolymers, lamellar $L_\beta$ materials,
biological membranes and tilted phases of Langmuir-Blodgett films
\cite{Mowald}. The origin of the tilt in the Sm{\it C} phase has
long been  the subject of some controversy, and several different
mechanisms of the Sm{\it A}--Sm{\it C} transition have been
proposed.

Elementary pictures of the tilt transition assume almost perfect
translational and orientational order. The transition here
corresponds to a collective tilt of the orientationally ordered
molecules. The closest analogy is a structural transformation, and
the nature of the ordering in the two phases is essentially the same
\cite{Goosens}. The tilt has a geometrical rather than statistical
definition, and the layer spacing $d$ decreases with tilt as $\cos
\Theta$. Recently, however, materials have been discovered with
diverse molecular structures
\cite{Naciri95,Radcliffe99,Giess99,Takanishi}, in which  the Sm{\it
C} layer spacing  is virtually constant.

A few such materials have long been known, but  regarded as
exceptional.  A qualitative explanation, due to de Vries
\cite{DeVries1},   requires that the Sm{\it A} molecules are not
oriented  \textit{along} the layer normal, but rather on a cone
surface \textit{around} it. The transition  now involves an ordering
of the molecular azimuthal angles, which results in a macroscopic
average tilt. This model is inconsistent with high orientational
order. Recent experiments do indeed indicate that the nematic order
parameter is abnormally low in de Vries smectics
\cite{revFrankJan,Giesselmann02,Collins00,huang}.

However, different materials exhibit various degrees of layer contraction
in the Sm{\it C} phase. Neither of the extreme limits above
is correct for real systems. Indeed  the accumulated experimental data permit
a study of a continuous range of materials between de
Vries-type and normal smectics. The problem of why some materials do not
show a layer contraction and why others do, is of considerable importance.
The underlying physical mechanism is both of fundamental physical interest
and of key practical importance. Chiral
tilted smectic ferro-, antiferro- and helielectric materials are
extremely promising for the new generation of fast electro-optic
displays as well as various non-display applications
\cite{DeGennes, Lagerwall}. Layer contraction causes the emergence of
chevron structures and 'zig-zag' defects, which present serious obstacles
for the commercialization of such devices.
However, novel de Vries-type materials are believed to
overcome this problem \cite{revFrankJan}.

In this letter we show that the de Vries-type Sm{\it
A}--Sm{\it C} phase transition can be successfully modeled both
phenomenologically and using a mean-field molecular approach.
The starting point is  the classical Chen-Lubensky phenomenological model
\cite{lubensky} for the Sm{\it A}--Sm{\it C} transition.
The smectic free energy is expressed
in terms of the wave vector ${\bf k}$ of the smectic density wave,
$k =2\pi/d$. We choose the $z$-axis along the director $\bf n$,
while the $x$-axis specifies the direction of possible tilt in the
smectic plane, $k_x=0$ in the Sm{\it A} phase. Neglecting fluctuations,
the free energy density is written as:
\begin{equation}\label{ChenLub}
\Delta F = D_{\parallel} (k_z^2-k_{\parallel}^2)^2 + C_{\bot} k_x^2
+ D_{\bot} k_x^4.
\end{equation}

The first term favors condensation of the smectic density wave at
$k_z=k_{\parallel}$. The last two terms describe the Sm{\it
A}--Sm{\it C} transition, i.e., the appearance of the nonzero
component $k_x$ when the coefficient $C_{\bot}$ changes sign. In the
Sm{\it C} phase the tilt angle $\Theta$ is given by $\tan \Theta
=k_x/k_z=-C_{\bot}/2D_{\bot} k_{\parallel}$. In this theory, $k_z
=k_{\parallel} \approx const$, yielding a  SmC layer spacing $d_c =
2\pi/k_c \propto \cos \Theta$. This is the  conventional Sm{\it C}
layer contraction, which  occurs mathematically because the
variables $k_z$ and $k_x$ in Eq.(\ref{ChenLub}) are uncoupled.

This result changes dramatically if one adds a simple coupling term $Ak^2_x
k^2_z$ to the model free energy (\ref{ChenLub}). This term is always
allowed by symmetry and is of the same order as the quadratic term
$C_{\bot} k_x^2$ because $k_z$ is not small. Now the total wave
vector of the smectic C structure is
\begin{equation}\label{kclub}
k_C^2 = k_{\parallel}^2 - {C_{\bot}(1-A/2D_{\bot})}D_\text{
eff}^{-1},
\end{equation}
where $D_\text{eff} =D_{\parallel} + D_{\bot}
-(1/2)(1-A/2D_{\bot})^2$. Moreover, the Sm{\it C} layer spacing
should  be constant for $A=2D_{\bot}$;  strictly speaking the
dependence $k_{\parallel}(T)$ yields a weakly temperature dependent
$k_C$. For intermediate values of $A$ between 0 and $2D_{\bot}$,
Eq.(\ref{kclub}) describes a slow contraction of smectic layers in
the Sm{\it C} phase, which is observed for many smectic materials.
Thus this generalized Chen-Lubensky model can provide a qualitative
description of layer contraction in both conventional and de Vries
smectics. But the picture is too general to describe specific
details of the Sm{\it A}--Sm{\it C} transition.

Further phenomenological progress requires the inclusion of the
experimental observation that the nematic order parameter $S$ is
abnormally small in de Vries materials. The nematic tensor order
parameter $Q_{ij}\approx S(n_i n_j-\delta_{ij}/3)$, neglecting the
small smectic biaxiality. The free energy can be expanded in powers
of $\bf Q$. Retaining linear and quadratic terms in ${\bf Q}$, the
model free energy is now:
\begin{multline}\label{phenomenol}
F = F_0 (S)- b_1 S^2 k^2 - e_1 ({\bf k\cdot\bf Q}
\cdot {\bf k}) \\
+ g_1 ({\bf k} \cdot {\bf Q} \cdot {\bf Q}
\cdot {\bf k})
+ b_2 S^2 k^4 + e_2 ({\bf k} \cdot {\bf Q}
\cdot {\bf k})^2\\
+ g_2 k^2({\bf k} \cdot {\bf Q}
\cdot {\bf k}) +c k^2({\bf k} \cdot {\bf Q} \cdot {\bf Q}
\cdot {\bf k}).
\end{multline}
Substituting the expression for ${\bf Q}$, we can rewrite the free
energy in terms of $k_z=({\bf k} \cdot {\bf n})$, $k$ and $S$. For
$g_2=c=0$ minimizing (\ref{phenomenol}) yields  $k^2_C \equiv const$.
In this case, no matter what the values of  the other model parameters, there is no
layer contraction in the Sm{\it C} phase.

In the case of nonzero $c$ or $g_2$ there exists a partial layer
contraction. In particular, minimizing the free energy
(\ref{phenomenol}) yields simple expressions for $k$ when
$g_1=g_2=0$:
\begin{equation}\label{kcphenom}
k_C^2 = k_0^2 \left( 1- c'(e'_1 S^{-1} -c')(2e'_2 -(c')^2)^{-1}
\right)
\end{equation}
and
\begin{equation}\label{thetaphenom}
\sin^2 \Theta = \frac{2}{3} \frac{2e'_2
+c'-e'_1S^{-1}(1+c')}{2e'_2-c'e'_1S^{-1}},
\end{equation}
where $k_0^2 = b'_1/2b'_2; e'_1=2e_1/3b_1; e'_2=2e_2/9b'_2;
c'=c/9b_1;$ and $b'_2 =b_2 -4c/27$. In the Sm{\it A} phase the wave
vector $k_A$ is expressed as:
\begin{equation}\label{kaphenom}
k_A^2 = k_0^2\left( 1+e'_1 S^{-1} \right)\left(1+2e'_2
+2c'\right)^{-1}.
\end{equation}

In the model (\ref{phenomenol}) the Sm{\it A}-Sm{\it C} transition
is governed by the dependence $S(T)$; of necessity $S$ is here far
from saturation. The transition occurs when $S$ reaches the critical
value $S_{AC} = e'_1(1+c')/(2e'_2 +c')$. In the Sm{\it A} phase (see
Eq.(\ref{kaphenom})) the layer spacing  always increases for
decreasing temperature; this is also true experimentally for all de
Vries materials \cite{revFrankJan}. In the Sm{\it C} phase the layer
contraction is controlled by the parameter $c'$. The spacing is
constant for $c'=0$, which can be regarded as ideal de Vries
behavior. On the other hand, $k_z=const$ when $c=e'_2 -1$. In this
case  the Sm{\it C} layer contraction is determined by the factor of
$\cos \Theta$; this is  ideal conventional smectic behavior.

This simple phenomenological model thus describes both limiting
cases, de Vries and conventional behavior. Furthermore, intermediate
cases observed in experiment correspond to intermediate values of
$c'$ between 0 and $e'_2 -1$. Our simple model expressions allow
excellent fitting of experimental data for absolutely different
materials of both de Vries and conventional type as illustrated by
Fig.1. \cite{odparam}

\begin{figure}
\centering\includegraphics[width=8cm]{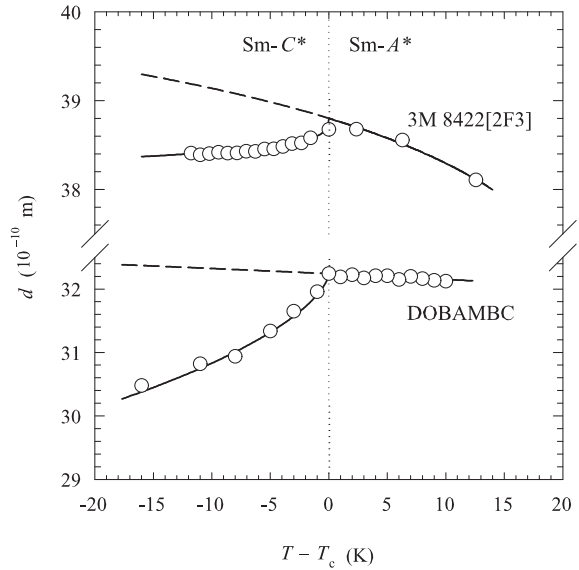}
\caption{Experimental data on variation of layer spacing in de
Vries-type 3M8422 (above) and conventional DOBAMBC (below) materials
fitted by Eqs.(4-6) with $e_2=0.18$ and $e_1=0.248,\ c=0.09,\
k_0=0.169{\text \AA}^{-1}$ for 3MB8422 and $e_1=0.439,\ c=0.49,\
k_0=0.238{\text\AA}^{-1}$ for DOBAMBC.}\label{Fig.1}
\end{figure}

The phenomenological model (\ref{phenomenol}) uses a free energy expansion in
terms of the nematic order parameter and components of the smectic
wave vector, which in general are  not small. Furthermore, the physical origin of the phenomenon is not clear.
To validate the model we develop a corresponding
molecular field theory, which does not require these
approximations.

Molecular models for the Sm{\it C} phase have been proposed by a
number of authors \cite{McMillan,Wulf,Vertogen,Goosens,sluckin1,Madhusudana}. Most of
the theories assume  the long molecular axis to be very highly oriented.
An exception is the recent paper by Govind and Madhusudana
\cite{Madhusudana}, which, however, focuses on a particular intermolecular
interaction which does not address de Vries materials. We believe that such
approaches cannot be used to describe de Vries-type smectics, where
the order parameter $S$ is relatively small. Below we present a brief
description of our general mean-field theory for the Sm{\it
A}-Sm{\it C} transition in the system of uniaxial molecules and
apply it to de Vries smectics.

We specify the relative position and orientation of two rigid
uniaxial molecules '1' and '2' by the intermolecular vector ${\bf
R}=R \ \bf{\hat r}$ and the molecular long axis unit vectors
${\bf a}_1$ and ${\bf a}_2$. For nonpolar molecules, the pair interaction
potential must be even in ${\bf a}_1$ and ${\bf a}_2$. For nonchiral
molecules, the potential is also even in ${\bf R}$. The pair
potential $U({\bf a}_1, {\bf R}, {\bf a}_2)$ can now be written as:
\begin{multline} \label{U12}
U({ {\bf a}_1, {\bf R}, {\bf a}_2})=u_1(R) \left[({\bf a}_1 \cdot
{\bf\hat r})^2+({\bf a}_2 \cdot {\bf \hat r})^2\right] + u_2(R)
({\bf a}_1\cdot {\bf a}_2)^2\\+u_3(R)({\bf a}_1\cdot{\bf a}_2)({\bf
a}_1\cdot{\bf \hat r})({\bf a}_2\cdot{\bf \hat r}) + u_4(R)({\bf
a}_1\cdot{\bf \hat r})^2({\bf a}_2\cdot{ \bf \hat r})^2,
\end{multline}
where all the possible terms quadratic in ${\bf a}_{1,2}$ have been
taken into account.

To construct a mean-field smectic free energy functional, we neglect
departures from perfect  smectic translational order and inter-layer interactions.
The free energy functional is now:
\begin{multline}\label{FMF}
F=\frac{\rho^2}{2}\int d{\bf a}_1 d{\bf a}_2 \ f_1({\bf a}) \ f_1
({\bf a}_2) \ U({ {\bf a}_1, {\bf r}, {\bf a}_2})\\
+\rho kT \int d{\bf a} \ f_1({\bf a}) \ln(f_1({\bf a})),
\end{multline}
where $\rho$ is the molecular number density per unit area of the
layer, and $f_1 ({\bf a})$ is the orientational distribution
function. Minimizing the free energy (\ref{FMF}) yields:
\begin{equation}\label{fMF}
f_1({\bf a})=Z^{-1}\exp\left[ -\rho U_{MF}({\bf a})/kT\right],
\end{equation}
where the mean field potential $U_{MF}({\bf a})$ is given by:
\begin{equation}\label{UMF}
U_{MF}({\bf a})= \int d{\bf a}_2 f_1({\bf a}_2)\int d{\bf R} \ U(
{\bf a}, {\bf R}, {\bf a}_2),
\end{equation}
and $Z=\int d{\bf a} \exp\left[ -\rho U_{MF}({\bf a})/kT\right]$ is
a normalization factor.

Substituting Eq.(\ref{U12}) into Eq.(\ref{UMF}) yields an explicit
expression for the mean field potential:
\begin{multline}\label{Uorder}
U_{MF}({\bf a})= w_1 P_2(\cos \gamma)+w_2 S_k P_2(\cos \gamma) \\
+w_3 P_k \sin^2 \gamma \cos 2\phi + w_4 C \sin 2\gamma \cos \phi .
\end{multline}
The angles $\gamma$ and $\phi$ are polar and azimuthal angles
specifying the orientation of the unit vector ${\bf a}$, i.e. ${\bf
a}=(\sin \gamma \cos \phi, \sin \gamma \sin \phi, \cos \gamma)$.
There are three orientational order parameters: $S_k =<P_2(\cos
\gamma)>$, $P_k=<\sin^2 \gamma \cos 2\phi>$ and $C=<\sin 2\gamma
\cos \phi>$, which determine completely the nematic tensor $\bf Q$.
The interaction constants are given by linear combinations of
integrals over the potential (\ref{U12}):
$w_1=-\bar{u_1}/3-(\bar{u}_3+\bar{u}_4)/9,
w_2=2\bar{u_2}/3+(\bar{u_3}+\bar{u_4})/9,
w_3=\bar{u_2}/2+\bar{u_3}/4+\bar{u_4}/8,
w_4=\bar{u_2}/2+\bar{u_3}/8$, with $ \bar{u}_\alpha=\int dR\ R\
u_\alpha(R)$.

The conventional order parameters of the Sm{\it C} phase are the
nematic order parameter $S$, the nematic tensor biaxiality $P$ and
the tilt angle $\Theta$. These can be expressed in terms of the parameters
$S_k, P_k$ and $C$:
\begin{eqnarray}
\tan 2\Theta =C(S_k -0.5 P_k)^{-1},\label{S}\\
S=S_k/4 + 3P_k/8+3C(4 \sin 2\Theta)^{-1},\label{P}\\
P=S_k/2 + 3P_k/4-C(2 \sin 2\Theta)^{-1} \label{C}.
\end{eqnarray}
In this theory the order parameter $C$, which  is proportional to the tilt
angle $\Theta$, is the primary order parameter at
the Sm{\it A}--Sm{\it C} transition. The biaxiality $P_k$ is
a secondary order parameter induced by the Sm{\it C} tilt.
The transition is thus a true order-disorder phase
transition. The low temperature phase is now characterized not only by a
tilt but by a non-zero order parameter $C$ explicitly defined as a statistical average.
\begin{figure}
\centering\includegraphics[width=8cm]{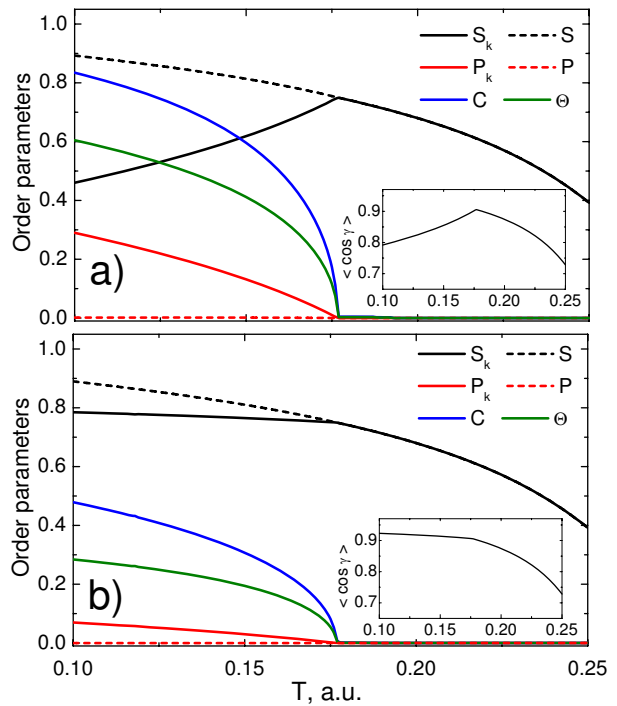}
\caption{Temperature dependence of order parameters for
'conventional' (a) and de Vries-type (b)  transitions. Inserts show
layer thickness changes. The model parameters are: $w_1=-0.05,\
w_2=-1,\ w_3=-0.9$ (a) or $w_3=-0.75$ (b), $w_4=-0.8$.}\label{Fig.2}
\vspace{-0.3cm}
\end{figure}

Relations (\ref{fMF}, \ref{Uorder}) enable the free energy
(\ref{FMF}) to be minimized self-conistently, which yields the
temperature dependences of parameters $S_k, P_k$, and $C$. The
conventional order parameters can then be established using Eqs.
(\ref{S}-\ref{C}). Remarkably, the resulting phase behavior has much
in common with that predicted phenomenologically. In both pictures
the Sm{\it A}--Sm{\it C} transition is directly driven by the growth
of $S$ with decreasing temperature. The onset of the Sm{\it C} phase
occurs at a critical value of the nematic order parameter, $S_{AC}=3
w_1/(4 w_4 - 3 w_2)$. Beyond this value, the tilt responds directly
to increases in orientational order: $\sin^2\Theta(T)\propto
[(S(T)-S_{AC})/S(T)]S_{AC}$.

We have selected the average $<|\cos \gamma|>$ as a simple surrogate
of the layer spacing \cite{franktheory}, i.e., we assume that the
spacing can be approximated by the average projection of the
molecules onto the layer normal. This assumption is supported by
recent experimental data \cite{Vij} which indicate that for several
different compounds  there exists a  good correlation between the
temperature variation of the smectic period $d$ and the order
parameter $S_k$. At the same time, $<|\cos \gamma|> \approx 1-0.5
<\sin^2 \gamma>$ and $S_k=1-1.5<\sin^2 \gamma>$, which yields
$<|\cos \gamma|> \approx (2+ S_k)/3$.

To describe  the transition in more detail we determine numerically
the free energy minima. The range of possible behaviors appears
to be surprisingly broad. A typical  conventional Sm{\it C} case is
shown in Fig.2a. The tilt angle is relatively large while the
biaxiality is small. In the Sm{\it C} phase itself the layer spacing
decreases approximately proportionally to $\cos \Theta$.
\begin{figure}
\centering\includegraphics[width=8cm]{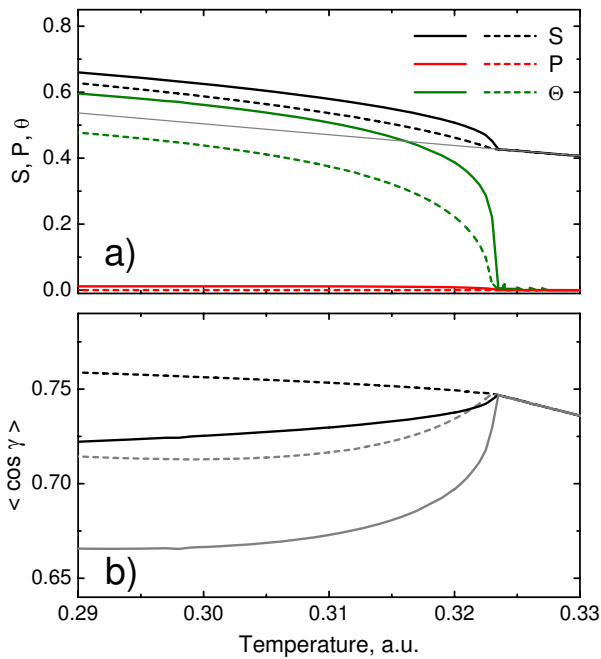} \caption{(a)
Effect of the transition on nematic order parameter calculated for
$w_1=-0.2,\ w_2=-1,\ w_3=-1.35$ (solid) or $w_3=-1.1$ (dashed), and
$w_4=-1.1$. The thin gray line shows the unperturbed values of $S$.
(b) Smectic layer spacing for the two cases above (black); layer
spacing calculated if  nematic order unperturbed
(gray).}\label{Fig.3} \vspace{-0.3cm}
\end{figure}

An example of a de Vries-type behavior is shown in Fig.2b. Here the
layer spacing is nearly constant in the Sm{\it C} phase and the tilt
angle is relatively small.  The two cases shown in Figs. 2a and 2b
differ only by changing the parameter $w_3$ in the model potential.
Thus, although the transition temperatures are the same, the
properties of the Sm{\it C} phase are qualitatively different. More
generally, we find that there exists an extended region in
parameter space for which the theory predicts smectics C with very
weak layer contraction.

The molecular model also predicts feedback effects in which the
presence of the tilt further increases the nematic orientation. This
can lead to an additional change of the layer spacing. The effect is
especially pronounced if the transition occurs at low $S_{AC}$ and
has recently been observed \cite{Kocot}. Typical behavior is shown
in Fig.3. Although the changes in the nematic order are relatively
weak, they trigger a qualitatively different temperature dependence
of layer thickness.

In the de Vries scenario of the Sm{\it A}-Sm{\it C}
transition,  the transition is governed by the temperature variation
of $S$. This is realistic only for relatively low nematic order.
In order for the picture to be valid,  an additional microscopic mechanism
is required to stabilize the smectic phases. In fact most de Vries materials
possess bulky siloxane or fluorinated
groups. These promote  microphase separation
\cite{revFrankJan,Dublin} which, in turn, favors smectic
ordering even in the absence of  nematic order. This mechanism dominates
in lyotropic lamellar phases where  layer formation
is primarily determined by water-amphiphiles  microphase separation.
In de Vries materials it appears that both microscopic mechanisms for smectic
ordering occur.

In conclusion, we have developed a generalized Chen-Lubensky
phenomenological model as well as a molecular model which describe
both conventional (layer-contracting) and de Vries smectics C as two
limiting cases. The theory also explains various intermediate cases.
These correspond to a crossover between  structural and
order-disorder Sm{\it A}--Sm{\it C} transitions, and  reflect the
observed properties of various real materials. The authors are
gratefull to P.Collings, C.C.Huang and Yu.Panarin for valuable
discussions, and to EPSRC (UK) for funding.


\begin{thebibliography}{99}
\bibitem{odparam}
The dependence $S(T)$ is approximated as $S_A(T) \propto
(T_A-T)^{\beta_A}$ in Sm$A$ phase and $(S_C(T)-S_{AC}) \propto
(T_C-T)^{\beta_C}$ in Sm$C$, which is justified by the observed
rapid growth of birefringence in the SmC phase of de Vries materials
\cite{revFrankJan} as well as by the molecular theory (see Fig.4).
\end{thebibliography}
\end{document}